\documentclass[12pt,twoside,fleqn]{article}

\usepackage{graphicx}

\newcommand{\figcaptionwidth}{15.cm}
\newcommand{\tabcaptionwidth}{13.75cm}

\newcommand{\be}{\begin{equation}}
\newcommand{\ee}{\end{equation}}

\newcommand{\bea}{\begin{eqnarray}}
\newcommand{\eea}{\end{eqnarray}}

\newcommand{\benn}{\begin{displaymath}}
\newcommand{\eenn}{\end{displaymath}}

\newcommand{\beann}{\begin{eqnarray*}}
\newcommand{\eeann}{\end{eqnarray*}}

\newcommand{\befig}{\begin{figure}}
\newcommand{\efig}{\end{figure}}

\newcommand{\fm}{\mbox{fm}}
\newcommand{\MeV}{\mbox{MeV}}
\newcommand{\GeV}{\mbox{GeV}}

\begin{document}


\title{
\vspace*{-1cm}
{\normalsize\rightline{HD-THEP-01-13}}
{\normalsize\rightline{CERN-TH/2001-080}}
{\normalsize\rightline{hep-ph/0103044}}
\vspace*{1.cm}
{\Large\bf 
Hard Thermal Photon Production in\\ 
Relativistic Heavy Ion Collisions}}
\author{}
\date{}
\maketitle

\vspace*{-2.cm}

\begin{center}

\renewcommand{\thefootnote}{\alph{footnote}}

{\large
  Frank~D.~Steffen$^{1,}$\footnote{Frank.D.Steffen@thphys.uni-heidelberg.de} and
  Markus~H.~Thoma$^{2,}$\footnote{Markus.Thoma@cern.ch}$^{,}$\footnote{Heisenberg
    fellow}}

\vspace*{0.5cm}

{\it $^1$Institut f\"ur Theoretische Physik, Universit\"at Heidelberg,\\
Philosophenweg 16, 69120 Heidelberg, Germany}

\smallskip

{\it $^2$Theory Division, CERN, CH-1211 Geneva 23, Switzerland}

\end{center}

\bigskip


\begin{abstract}
  
  The recent status of hard thermal photon production in relativistic heavy ion
  collisions is reviewed and the current rates are presented with emphasis on
  corrected bremsstrahlung processes in the quark-gluon plasma (QGP) and
  quark-hadron duality. Employing Bjorken hydrodynamics with an EOS supporting
  the phase transition from QGP to hot hadron gas (HHG), thermal photon spectra
  are computed. For SPS 158 GeV Pb+Pb collisions, comparison with other
  theoretical results and the WA98 direct photon data indicates significant
  contributions due to prompt photons. Extrapolating the presented approach to
  RHIC and LHC experiments, predictions of the thermal photon spectrum show a
  QGP outshining the HHG in the high-$p_T$-region.

\vspace{1.cm}


\medskip


\end{abstract}

%
\newpage
\renewcommand{\thefootnote}{\arabic{footnote}}
\setcounter{footnote}{0}
%
\section{Introduction}
\label{Introduction}
%
Hard real photons are, as dileptons, promising for providing a signature of a
quark-gluon plasma (QGP) possibly produced in relativistic heavy-ion
collisions~\cite{RUUSKANEN_1992}. Direct photon spectra have been measured by
the WA80 (upper limit) and the WA98 collaboration at the
SPS~\cite{WA80_1996,WA98_2000} and further data is expected from the RHIC and
LHC heavy-ion experiments~\cite{PHENIX_1998,ALICE_1995_96}. For theoretical
investigations on the SPS direct photon data and on RHIC and LHC predictions,
prompt and thermal photon production rates must be convoluted with the
space-time evolution of the fireball, while photons from the decay of hadrons
after freeze-out are already subtracted in the experimental analysis.

In this work, the present status of hard thermal photon production in a QGP and
a hot hadron gas (HHG) is reviewed critically. Contributions from the QGP are
presented with emphasis on bremsstrahlung processes that are illustrated for the
first time in their corrected form. Next, the photon producing processes in the
HHG are discussed and a conservative estimate of the thermal rate is given. The
review of the rates is completed by addressing the question of quark-hadron
duality.  With the obtained insights, the recent estimates of the rates are
employed together with the well-understood Bjorken
hydrodynamics~\cite{BJORKEN_1983,GYULASSY_1984} to extract the essential
features of the thermal and prompt photon yield in the phase transition
scenario~\cite{SATZ_1994}. Comparison with other theoretical work demonstrates
the competence of the simple hydrodynamical model at SPS energies and
substantiates subsequent inspections of the WA98 data on direct photon
production in SPS $158\;\GeV$ Pb+Pb collisions. For RHIC and LHC experiments,
the rate estimations with the same hydrodynamical model are used to predict the
$p_T$-range in which the QGP outshines the HHG.
%
\section{Thermal Photon Production in the QGP}
\label{Thermal_Photon_Production_QGP}
%
The production rate for hard ($E \gg T$) thermal photons\footnote{In the
  experimentally interesting case $E\gg T$ analytic expressions can be derived
  since e.g.\ Boltzmann distributions can be used for the outgoing
  partons~\cite{KAPUSTA_1991}.} from an equilibrated QGP has been calculated in
perturbative thermal QCD applying the hard thermal loop (HTL)
resummation~\cite{BRAATEN_1990} to account for medium effects. The {\em Compton
  scattering} and {\em $q\bar{q}$-annihilation} contribution
\be
        \left. E\,\frac{dN}{d^4x\,d^3p} \,\right|_{1-loop} = 
        0.0281\,\alpha \alpha_s \ln \left(\frac{0.23\,E}{\alpha_s\,T}\right)
        \,T^2\,e^{-E/T},
\label{1-loop}
\ee
is derived from the 1-loop HTL photon-polarisation
tensor~\cite{KAPUSTA_1991,BAIER_1991,TRAXLER_1995}, while the contributions from
{\em bremsstrahlung},
\be
        \left. E\,\frac{dN}{d^4x\,d^3p} \,\right|_{bremss} = 
        0.0219\,\alpha \alpha_s
        \,T^2\,e^{-E/T},
\label{bremss}
\ee
and {\em $q\bar{q}$-annihilation with an additional scattering in the medium},
\be
        \left. E\,\frac{dN}{d^4x\,d^3p} \,\right|_{q\bar{q}-aws} =  
        0.0105\,\alpha \alpha_s
        \,E\,T\,e^{-E/T},
\label{qqbar-aws}
\ee
are obtained from the 2-loop HTL photon-polarisation
tensor~\cite{AURENCHE_1998}, where all three rates are listed for a two-flavored
($N_f = 2$) QGP.  Surprisingly, the 2-loop rates, (\ref{bremss})
and~(\ref{qqbar-aws}), show up at order~$\alpha\alpha_s$ and enhance the
spectrum from the QGP phase by about a factor of 3 in the experimentally
relevant $p_T$-range. Since earlier investigations on the importance of
bremsstrahlung processes in the
QGP~\cite{SRIVASTAVA_1999,SRIVASTAVA_SINHA_1999,STEFFEN_1999} employed the
rates~(\ref{bremss}) and~(\ref{qqbar-aws}) multiplied {\em erroneously} by a
factor of 4, we present this behavior for the first time in its corrected
form.\footnote{Due to a miscalculation of the two $N_f$-dependent constants
  $J_T$ and $J_L$ (exactly a factor 4 too large)
  in~\cite{AURENCHE_1998,SRIVASTAVA_1999}, the derivation of the rates led to
  Eqs.~(\ref{bremss}) and~(\ref{qqbar-aws}) both multiplied by a factor of 4. In
  this way, an {\em erroneous} enhancement of the spectrum from the QGP phase by
  about {\em one order of magnitude} was found in the experimentally relevant
  $p_T$-range~\cite{SRIVASTAVA_1999,SRIVASTAVA_SINHA_1999,STEFFEN_1999}.}  The
enhancement is anticipated by comparing the rates for fixed temperatures as
shown in Fig.~1 and is documented in Fig.~2, where the thermal photon spectrum
from the QGP, calculated in the model discussed below, is presented.\footnote{In
  the QGP rates, $\alpha_s(T)=\frac{6\pi}{(33 - 2
    N_f)\ln(8T/T_c)}$~\cite{KARSCH_1988} is applied, where the number of flavors
  present in the QGP is set to $N_f=2$ and the transition temperature is set to
  $T_c=170\;\MeV$.} Traced back to strong collinear
singularities~\cite{AURENCHE_1998}, this 2-loop enhancement substantially
increased the interest in higher loop contributions. In fact, for real photons
the 3-loop contribution turns out to be of the same order in $\alpha_s$ as the
2-loop contribution~\cite{AURENCHE_2000}. Since this is very likely the case for
higher loop contributions as well, one can conclude that thermal photon
production in the QGP is a non-perturbative mechanism that cannot be accessed in
perturbative HTL resummed thermal field theory. Interestingly, ascribing the
quarks a finite mean free path in the QGP, which simulates the
Landau-Pomeranchuk-Migdal (LPM) effect, does not eliminate important
non-perturbative aspects but helps to disarm the collinear
singularities~\cite{AURENCHE_LPM_2000}. This points at possible destructive
interferences via the LPM effect that do not allow one to interpret the sum of
the rates~(\ref{1-loop}), (\ref{bremss}), and~(\ref{qqbar-aws}) as a lower
limit. Adding further uncertainties such as the $g \ll 1$ assumption in the HTL
calculations contrasted with realistic values of the strong coupling, $g = 2 -
3$, one must consider the sum of the rates~(\ref{1-loop}), (\ref{bremss}),
and~(\ref{qqbar-aws}) only an educated guess. However, it is employed in this
work as the best result available. Extensions to a QGP not in chemical
equilibrium, which seems realistic at RHIC and LHC, can also be
found~\cite{STRICKLAND_1994,TRAXLER_1996,SRIVASTAVA_1997,BAIER_1997,MUSTAFA_CNEQ_2000}.
\begin{figure}
  \includegraphics[width=13.5cm]{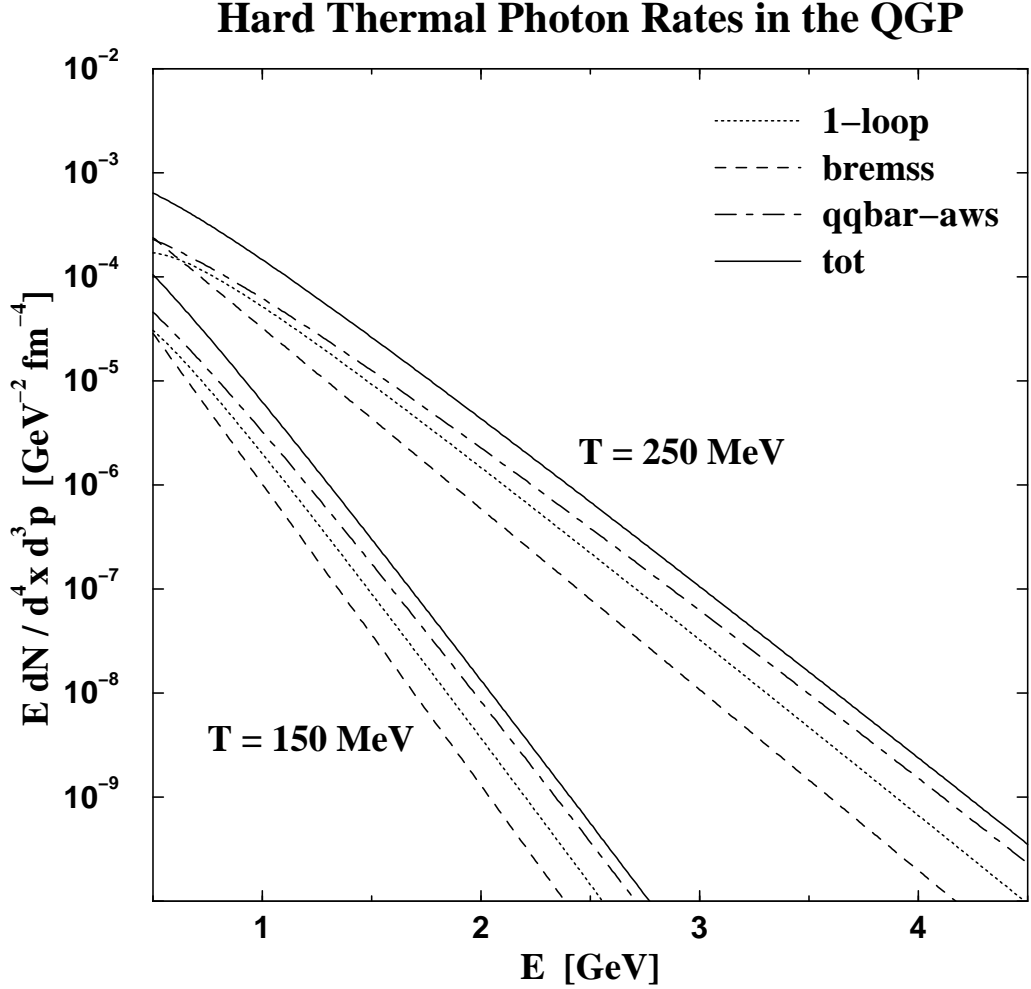}
  \parbox{\figcaptionwidth}{\caption[Hard Thermal Photon Rates in the QGP] {Hard
      Thermal Photon Rates in the QGP. For two fixed temperatures, $T=150\;\MeV$
      (lower lines) and $T=250\;\MeV$ (upper lines), the QGP hard thermal photon
      rates~(\ref{1-loop}), (\ref{bremss}), and (\ref{qqbar-aws}) are
      illustrated in the dotted, dashed and dot-dashed lines respectively, where
      the $\alpha_s(T)$-parameterization of Karsch~\cite{KARSCH_1988} is
      applied. The sum indicating the total QGP contribution up to 2-loop order
      is displayed in the solid lines. Inclusion of the 2-loop processes
      enhances the total QGP rate by about a factor of 3.}}
\label{Fig_hard_thermal_photon_rates_QGP}
\end{figure} 
\begin{figure}
  \includegraphics[width=13.5cm]{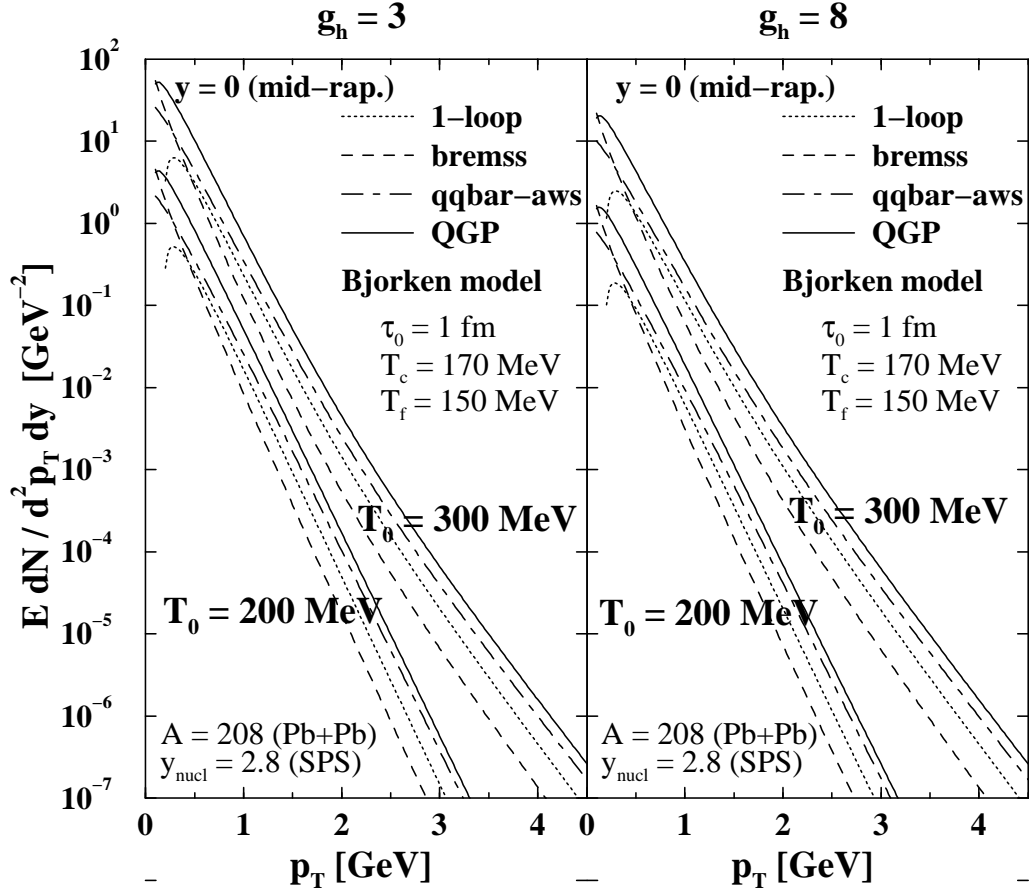}
  \parbox{\figcaptionwidth}{\caption[Hard Thermal Photon Spectrum from the QGP]
    {Hard Thermal Photon Spectrum from the QGP. The QGP contributions to the
      hard thermal photon yield from the rates~(\ref{1-loop}), (\ref{bremss}),
      (\ref{qqbar-aws}), and the sum of these are illustrated in the dotted,
      dashed, dot-dashed and solid lines respectively. The integration of the
      rates over the space-time evolution of the fireball was performed in
      Bjorken hydrodynamics described in the text. Results are shown for two
      different initial temperatures, $T_0 = 200\;\MeV$ (lower lines) and $T_0=
      300\;\MeV$ (upper lines), and two different hadron gas EOS's characterized
      by $g_h = 3$ (left plot) and $g_h = 8$ (right plot) with the remaining
      parameters set to the displayed {\em typical} values. The 2-loop processes
      enhance the total QGP spectrum by about a factor of 3.}}
\label{Fig_thermal_photon_spectrum_QGP}
\end{figure} 
%
%
\section{Thermal Photon Production in the HHG}
\label{Thermal_Photon_Production_HHG}
%
The thermal photon production in an equilibrated hot hadron gas (HHG) is
computed in effective theories with hadronic couplings inferred from experiment.
A first investigation based on a model Lagrangian describing the interaction of
$\pi$, $\rho$, and $\eta$ mesons with photons found dominant contributions from
the reactions $\pi \pi \rightarrow \rho \gamma$ and $\pi \rho \rightarrow \pi
\gamma$ and from the decays $\omega \rightarrow \pi \gamma$ and $\rho
\rightarrow \pi \pi \gamma$~\cite{KAPUSTA_1991,NADEAU_1992}.  By considering
additionally the $\pi \rho \rightarrow a_1 \rightarrow \pi \gamma$ reaction, a
strong enhancement of the rate was observed~\cite{XIONG_1992}. This triggered a
more complete and consistent computation using an effective chiral Lagrangian
with $\pi$, $\rho$, and $a_1$ mesons. The result was an even higher rate due to
the presence of the $a_1$ meson also in the $\pi \pi \rightarrow \rho \gamma$
and $\rho \rightarrow \pi \pi \gamma$ reactions~\cite{SONG_1993,SONG_1998}.
Keeping the strong dependence on the model Lagrangian and the uncertainties
associated with medium effects in mind, the result parameterized
in~\cite{SONG_1998} supplemented by the $\omega \rightarrow \pi \gamma$
decay~\cite{KAPUSTA_1991} can be considered as a {\em conservative} expression
for the emissivity of the HHG. It is employed in this work using the formulas
listed in the Appendix. For hard photons, $E > 1\;\GeV$, we found a rough
estimate of this sum by multiplying the parameterization given in Eq.~(18)
of~\cite{XIONG_1992} by a factor of two
\be
        \left. E\,\frac{dN}{d^4x\,d^3p} \,\right|_{had} = 
        4.8\,T^{2.15}\,e^{-1/(1.35\,T\,E)^{0.77}}\,e^{-E/T},
\label{Markus_suggestion}
\ee
where photon energy~$E$ and temperature~$T$ are to be given in GeV to obtain the
rate in units of $\fm^{-4}\GeV^{-2}$. Results from other Lagrangians can be
found in~\cite{KIM_1996,HALASZ_1998} and medium effects are investigated
in~\cite{SONG_1998,HALASZ_1998,STEELE_1996,SARKAR_1998}, where a significant
increase of the static rate is observed by dropping the in-medium meson masses
that is, however, in part compensated in the spectrum through a consequent
modification of the fireball evolution. Further, the influence of finite
chemical potential~\cite{STEELE_1996}, finite baryon density~\cite{STEELE_1997}
and additional reactions involving $a_1$, $b_1$, $K_1$, and other strange
mesons~\cite{HAGLIN_1994,LEE_1998} have been considered but are neglected in
this work.
%
\section{Quark-Hadron Duality}
\label{Quark-Hadron_Duality}
%
In order to test the hypothesis of quark-hadron duality in photon
production~\cite{RAPP_1999,GALLMEISTER_2000}, we compare the discussed QGP and
HHG hard thermal photon rates for two fixed temperatures, $T = 150$ and
$200\;\MeV$, as shown in Fig.~3, where the solid and dashed lines indicate the
QGP and the HHG contributions respectively.
\begin{figure}
  \includegraphics[width=14.cm]{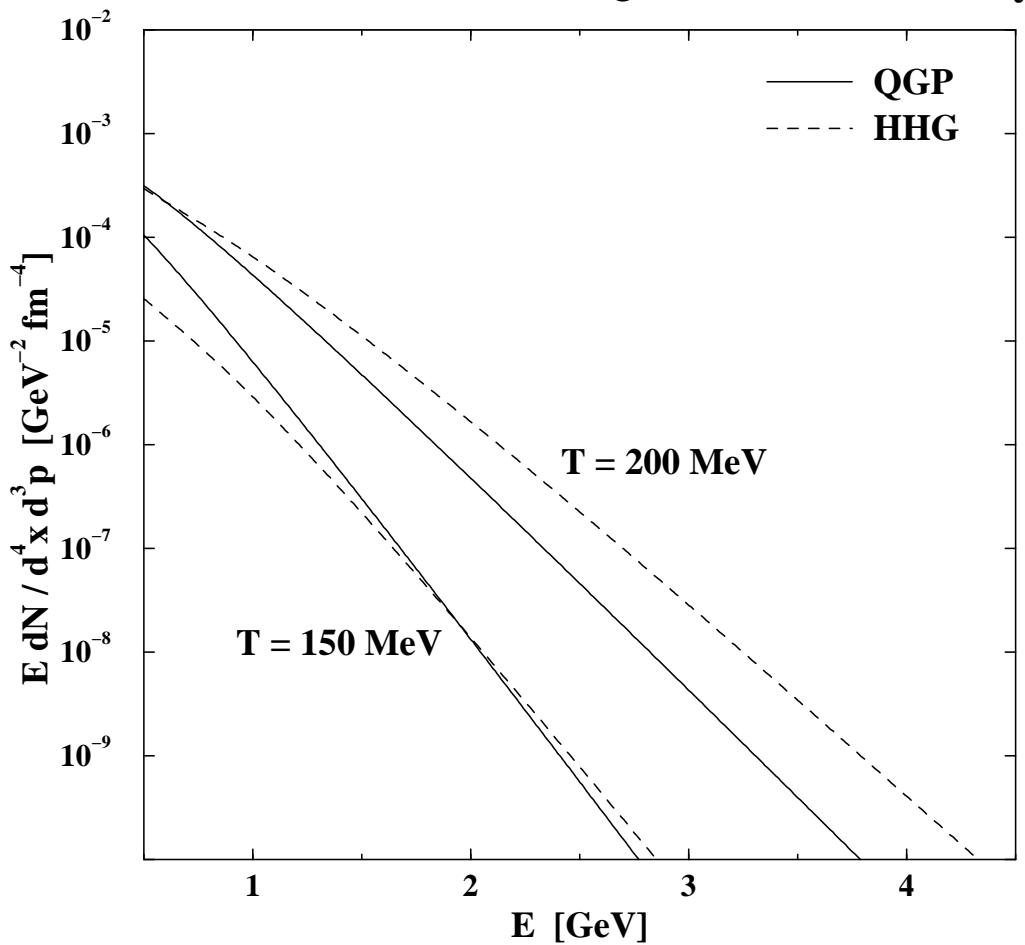}
  \parbox{\figcaptionwidth}{\caption[Quark-Hadron Duality in the Hard Thermal
    Photon Rates] {Quark-Hadron Duality in the Hard Thermal Photon Rates. The
      sum of the QGP contributions~(\ref{1-loop}), (\ref{bremss}),
      and~(\ref{qqbar-aws}), illustrated in the solid lines, and that of the HHG
      contributions (Song's results~\cite{SONG_1993,SONG_1998} supplemented by
      the $\omega \rightarrow \pi \gamma$ decay~\cite{KAPUSTA_1991}),
      illustrated in the dashed lines, are compared for two fixed temperatures,
      $T = 150\;\MeV$ (lower lines) and $T = 200\;\MeV$ (upper lines), where the
      $\alpha_s(T)$-parameterization of Karsch~\cite{KARSCH_1988} is applied in
      the QGP rates. Considering the uncertainties in the rates, quark-hadron
      duality in the rates at a certain temperature can neither be confirmed nor
      be ruled out.}}
\label{Fig_hard_thermal_photon_rates_duality}
\end{figure} 
For temperatures between $150\;\MeV$ and $200\;\MeV$, an energy window can be
found, where the steeper QGP and the flatter HHG emissivities cross. Considering
the uncertainties in the rates, the validity of the classic statement 'the
hadron gas shines as brightly as the quark-gluon plasma'~\cite{KAPUSTA_1991} can
neither be confirmed nor be ruled out, at least in a certain temperature regime,
by taking into account the 2-loop QGP processes and the $a_1$ meson in the HHG.
However, as indicated by Fig.~3, there is no reason to assume that the photon
production rates from the two phases coincide at all temperatures. But even if
the rates of the two phases agree in the relevant temperature regime, the
spectrum might be changed by the presence of the QGP due to a different
space-time evolution.
%
\section{Thermal Photon Spectrum}
\label{Thermal_Photon_Spectrum}
%
The observable quantity is the spectrum of direct photons which is calculated
theoretically by convoluting the prompt and thermal photon rates with the
space-time evolution of the fireball produced in heavy-ion collisions.
Concentrating on the effects of the thermal photon rates, we describe the
thermalized collision phase in
Bjorken-1+1-hydrodynamics~\cite{BJORKEN_1983,GYULASSY_1984} taking an EOS with a
first-order phase transition from a QGP, modelled by a massless two-flavored
parton gas, to a HHG, modelled by a massless pion gas. This well-understood,
simple approach allows us to demonstrate very clearly the influences of the
rates on the spectrum, and interestingly, comparison with other work shows it to
be as competent as 2+1 hydrodynamical models at least at SPS
energies~\cite{STEFFEN_1999}. This means that at SPS the transverse expansion
has only minor effects on the photon production.
\subsection{Comparison with other Work}
The WA98 direct photon data analysis of Gallmeister et al.\ obtained in a model
describing a {\em spherically} symmetric expansion~\cite{GALLMEISTER_2000} can
be reproduced in the simple model with the discussed rates.\footnote{The $\omega
  \rightarrow \pi \gamma$ decay is not taken into account since electromagnetic
  meson decays were subtracted in the experimental analysis of the WA98
  collaboration~\cite{PEITZMANN_2000_PC}.} By assuming a thermalization time of
$\tau_0 = 1\;\fm$ and equal initial, transition, and freeze-out temperatures of
$T_0 = T_c = T_f = 170\;\MeV$ as in~\cite{GALLMEISTER_2000}, i.e.\ only a mixed
phase, this single temperature value suffices to describe the WA98
data~\cite{WA80_1996,WA98_2000} when the prompt photon estimation
of~\cite{GALLMEISTER_2000} is added.

It is also possible to reproduce the WA98 direct photon data analysis of
Srivastava et al.~\cite{SRIVASTAVA_2000}, which does not necessitate prompt
photons but instead initial conditions that are rather extreme for SPS, i.e.\ a
very small thermalization time of $\tau_0 = 0.2\;\fm$ and a very high initial
temperature of $T_0 = 335\;\MeV$. Since their analysis was performed using the
2-loop QGP rates by mistake multiplied by a factor of 4, even smaller
thermalization times together with higher initial temperatures are demanded when
applying the corrected 2-loop QGP rates. This is due to the relative importance
of the QGP rates in the high $p_T$-range.  Concentrating on a validity-check of
the simple model, a comparison using {\em the same erroneous} rates is
performed. While the QGP thermal photon spectrum obtained in our simple model
matches directly the one of Srivastava et al.~\cite{SRIVASTAVA_2000}, we must
increase the effective degrees of freedom in the ideal massless pion gas from
the actual value of $g_h = 3$ to an effective one of $g_h =8$ in order to
achieve the fit in the HHG thermal photon spectrum. This points to the {\em
  rich} HHG EOS employed in~\cite{SRIVASTAVA_2000} having a much stronger effect
than transverse expansion.
\subsection{Thermal Photon Spectrum from the QGP}
These insights into the reliability of the simple model substantiate the QGP
thermal photon spectrum calculated in this model and shown in Fig.~2 to
illustrate the effect of 2-loop processes in the QGP. Displaying the spectrum
not only for $g_h = 3$ (left plot) but also for $g_h =8$ (right plot)
demonstrates additionally the effect of a richer HHG EOS, which lowers the QGP
thermal photon yield by reducing the lifetime of the mixed phase.
\subsection{Comparison with WA98 Direct Photon Data}
The total thermal photon spectrum emerging from the discussed rates in the
simple model can be exploited by using the WA98 measurements of the direct
photon yield~\cite{WA98_2000} to specify upper limits on the initial
temperature, $T_0^{max}$, reached in the SPS $158\;\GeV$ Pb+Pb collisions. This
is presented in Fig.~4.
\begin{figure}
  \includegraphics[width=14.cm]{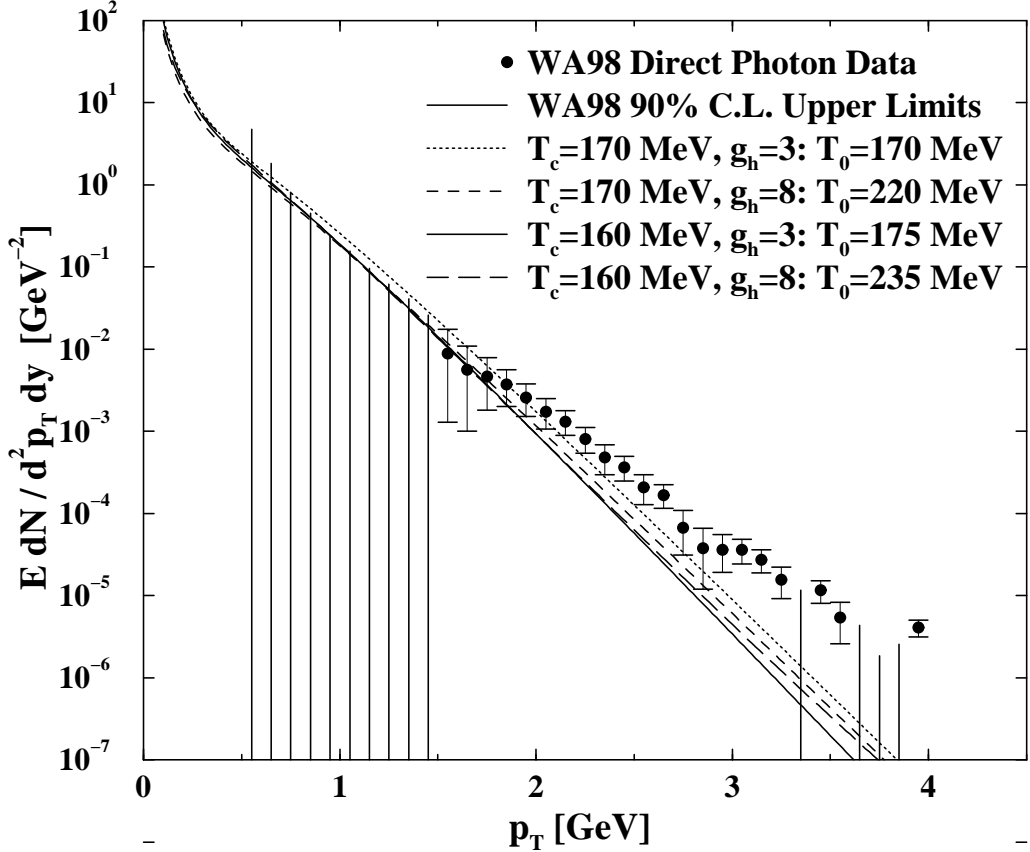}
  \parbox{\figcaptionwidth}{\caption[WA98 Direct Photon Data and Upper Limits on
    the Initial Temperature] {WA98 Direct Photon Data and Upper Limits on the
      Initial Temperature. The theoretical spectra supporting the phase
      transition from QGP to HHG (solid, short-dashed, and long-dashed lines)
      are compared with the experimental upper limits (vertical lines) and data
      (points with error bars). The discrepancy in the high-$p_T$-region
      demonstrates that a significant prompt photon contribution is necessary
      for a theoretical explanation of the experimental results within the phase
      transition scenario.}}
\label{Fig_WA98_data_T0_max}
\end{figure} 
For {\em typical} parameters 
$\tau_0 = 1\;\fm$, $T_c = 170\;\MeV$, $T_f = 150\;\MeV$, 
nucleon number $A=208$ (corresponding to Pb + Pb collisions), projectile
rapidity $y_{nucl}=2.8$ (corresponding to a center-of-mass energy of
$\sqrt{s}=17\;\GeV$), and with an ideal massless pion gas of $g_h = 3$ effective
degrees of freedom, even with an initial temperature of $T_0 = T_c = 170\;\MeV$
(dotted line) the computed spectrum {\em exceeds} the experimental limits at low
$p_T$! Simulating the richer HHG EOS by setting $g_h = 8$, the lifetime of the
mixed and pure HHG phase drops, resulting in a decrease of the spectrum that
brings back the possibility of having a phase transition scenario with
$T_0^{max}=220\;\MeV$ (short dashed line).  By reducing the critical temperature
to $T_c = 160\;\MeV$, the mean temperature is reduced enabling also for $g_h =
3$ the phase transition scenario with $T_0^{max}=175\;\MeV$ (solid line). For
$g_h = 8$ and $T_c = 160\;\MeV$, $T_0^{max}=235\;\MeV$ (long dashed line) is
found, which might be interpreted as an upper bound on the initial temperature
reached in the SPS $158\;\GeV$ Pb+Pb collisions.\footnote{Due to the mentioned
  possible destructive interferences via the LPM effect in the QGP, the
  presented $T_0^{max}$ values cannot represent {\em rigid} upper bounds. They
  rather indicate the magnitude and point to the {\em relative} differences
  caused by different model scenarios.}  However, the spectra supporting the
phase transition from QGP to HHG cannot describe the direct photon data for $p_T
> 2\;\GeV$ as shown in Fig.~4. This discrepancy in the high-$p_T$-region
demonstrates that a significant prompt photon contribution is necessary for a
theoretical explanation of the experimental results within the phase transition
scenario.
\subsection{Predictions for RHIC and LHC}
Finally, we employ the simple model to extract predictions for RHIC and LHC.
Although earlier comparisons with~\cite{SRIVASTAVA_1999} revealed a more
important transverse expansion~\cite{STEFFEN_1999}, it is interesting to locate
in the above approach the $p_T$-range in which the QGP outshines the HHG.
Assuming a three-flavored QGP, the prefactors on the rhs in Eqs.~(\ref{1-loop}),
(\ref{bremss}), and~(\ref{qqbar-aws}) become 0.0338, 0.0281, and 0.0135
respectively. Further, the SPS, RHIC, and LHC parameters
of~\cite{SRIVASTAVA_1999} are adopted as summarized in Tab.~1.
\begin{table}
\centering      
\begin{tabular}{|c|c|c|c|c|c|c|c|}
\hline
Accelerator / & Experiment & $A$   & $\sqrt{s}$    & $y_{nucl}$ & $dN/dy$ & $\tau_0$ & $T_0$ \\
Collider      &            &       & [A$\cdot$GeV] &            &         & [fm]     & [MeV] \\ 
\hline\hline
SPS           & WA98       & $208$ & 17            & 2.8        &  825    & 1.0      & 190   \\ 
\hline\hline
RHIC          & PHENIX     & $208$ & 200           & 5.3        & 1734    & 0.5      & 310   \\
LHC           & ALICE      & $208$ & 5500          & 8.6        & 5625    & 0.5      & 450   \\
\hline
\end{tabular}
\begin{center}
\parbox{\tabcaptionwidth}{
\caption[Parameters used in the Predictions of the Thermal Photon Spectrum for RHIC and LHC Experiments]{Parameters used in the Predictions of the Thermal Photon Spectrum for RHIC and LHC Experiments. The listed values for multiplicity $dN/dy$, thermalization time $\tau_0$, and initial temperature $T_0$ are the estimates given in~\cite{SRIVASTAVA_1999}.}}
\label{Tab_SPS_RHIC_LHC_parameters}
\end{center}
\end{table}
Figure~5 shows our results for the QGP, HHG, and sum of both contributions in
the dashed, dotted, and solid lines respectively.
\begin{figure}
  \includegraphics[width=13.75cm]{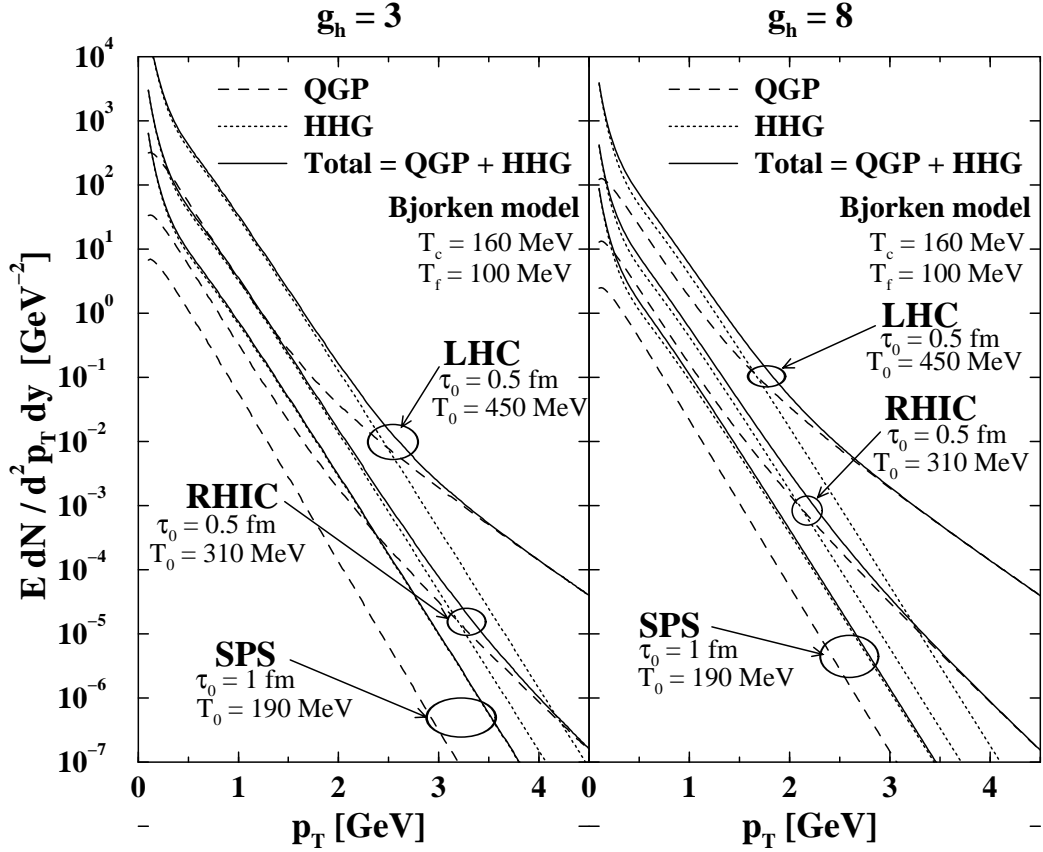}
  \parbox{\figcaptionwidth}{\caption[QGP vs.\ HHG Photon Emissivities in SPS,
    RHIC, and LHC Pb+Pb Collisions] {QGP vs.\ HHG Photon Emissivities in SPS,
      RHIC, and LHC Pb+Pb Collisions. For $g_h=3$ ($g_h=8$), the QGP outshines
      the HHG for $p_T > 3.2\;\GeV$ ($p_T > 2.2\;\GeV$) at RHIC and for $p_T >
      2.5\;\GeV$ ($p_T > 1.7\;\GeV$) at LHC, while at SPS photons from HHG
      dominate the spectrum for $g_h=3$ and $g_h=8$ at all $p_T$'s. The initial
      conditions vary from ($\tau_0 = 1\;\fm$, $T_0 = 190\;\MeV$) at SPS over
      ($\tau_0 = 0.5\;\fm$, $T_0 = 310\;\MeV$) at RHIC to ($\tau_0 = 0.5\;\fm$,
      $T_0 = 450\;\MeV$) at LHC, while the same values of $T_c = 160\;\MeV$ and
      $T_f = 100\;\MeV$ are chosen.}}
\label{Fig_SPS_RHIC_LHC}
\end{figure} 
The spectra for $g_h = 3$ are shown on the left and the ones for $g_h = 8$ on
the right. As a reference point, the SPS result obtained with the parameters
of~\cite{SRIVASTAVA_1999} is also presented. For $g_h=3$ ($g_h=8$), the QGP
outshines the HHG for $p_T > 3.2\;\GeV$ ($p_T > 2.2\;\GeV$) at RHIC and for $p_T
> 2.5\;\GeV$ ($p_T > 1.7\;\GeV$) at LHC, while at SPS photons from HHG dominate
the spectrum for $g_h=3$ and $g_h=8$ at all $p_T$'s. Since the employed simple
model does not describe deviations from chemical equilibrium, the reader is
referred to~\cite{MUSTAFA_CNEQ_2000}, where this issue is addressed
systematically and RHIC and LHC predictions are provided for the QGP photon
emissivity.
%
\newpage
\section{Conclusion}
\label{Conclusion}
%
We have discussed the hard thermal photon production in relativistic heavy ion
collisions using the most recent estimates for the photon production rates from
the QGP and the HHG. For the first time, the enhancement of the QGP thermal
photon rate and yield due to bremsstrahlung processes was presented in its
corrected form. The consideration of these processes hinted at the recently
confirmed non-perturbative nature of thermal photon production in the
QGP~\cite{AURENCHE_2000}. The derivation of a reliable rate requires now new
developments in thermal field theory~\cite{AURENCHE_LPM_2000}.  For thermal
photon production in the HHG, the strong dependence on assumptions regarding the
thermalized hadron species, their interactions, and the role of medium effects
was emphasized. We identified the best estimates available for the QGP and HHG
rates and presented a convenient parameterization for the total HHG rate. Using
these estimates, we could find no indication of a quark-hadron duality in the
photon production rate.

Integrating the rate estimations over the space-time evolution of the fireball
modelled in Bjorken hydrodynamics, we obtained the thermal photon spectrum.
Comparison with other models, which describe transverse expansion of the
fireball, demonstrated the competence of the simple model for the phase
transition scenario at SPS energies and substantiated the subsequent
investigation of the WA98 direct photon data.  We found that this experimental
data allows no conclusion to be drawn about the existence of a QGP phase at SPS
but can be explained by a conservative thermal source {\em plus} prompt photons.
Finally, the simple model was employed also for RHIC and LHC energies and
predicted a QGP outshining the HHG in the high-$p_T$-range. If this picture is
not spoiled by transverse expansion and chemical non-equilibrium or covered by
dominant prompt photon contributions, the RHIC and LHC experiments might see
thermal photons from the QGP in a certain photon momentum range that could
provide the desired signature of the QGP.

As RHIC is already taking data and LHC under construction, a reliable expression
for the thermal photon production rate in the QGP is of utmost importance.
Optimistically the final result could reveal photons to be a smoking gun for the
production of the QGP. In order to reduce the uncertainties concerning the
fireball evolution, we recommend a combined investigation of real photon,
dilepton, and hadron spectra, similar to the one of Sollfrank et
al.~\cite{SOLLFRANK_1997}, in which aspects such as different EOS's, finite
baryon density, chemical non-equilibrium, and transverse expansion should be
addressed in a systematic way. Once such an investigation is completed in
accordance with the SPS data, it should be extended to RHIC and LHC energies to
obtain serious predictions.
%
\section*{Acknowledgements}
\label{Acknowledgements}
%
We thank Munshi Mustafa for helpful and interesting discussions.
%
\begin{appendix}
\section*{Appendix}
\label{Appendix}
%
The employed rate for thermal photon production in the HHG is composed of the
exact expression for the decay $\omega \rightarrow \pi
\gamma$~\cite{KAPUSTA_1991} and the parameterizations for the processes $\pi \pi
\rightarrow \rho \gamma$, $\pi \rho \rightarrow \pi \gamma$, and $\rho
\rightarrow \pi \pi \gamma$, in which the $a_1$ meson is taken into account
properly~\cite{SONG_1993,SONG_1998}. For completeness and convenience, the
explicit formulas are listed. By inserting~$E$, $T$, and the pion mass~$m_{\pi}$
in GeV, the following parameterizations for $process$ = $\pi \pi \rightarrow
\rho \gamma$, $\pi \rho \rightarrow \pi \gamma$, and $\rho \rightarrow \pi \pi
\gamma$ reproduce the corresponding rates in units of
$\fm^{-4}\GeV^{-2}$~\cite{SONG_1998}
\bea
        \left. E\,\frac{dN}{d^4x\,d^3p} \,\right|_{process} & = & 
        T^2\,e^{-E/T}\,F_{process}(T/m_{\pi},E/m_{\pi}),
\eea
where
\bea
        F_{\pi\pi \rightarrow \rho\gamma}(x,y) 
        & = & \exp[-12.055 + 4.387x +(0.3755+0.00826x)y  
\nonumber\\      &   & + (-0.00777+0.000279x)y^2 + (5.7869-1.0258x)/y 
\nonumber\\      &   & + (-1.979+0.58x)/y^2],
\\
        F_{\pi\rho \rightarrow \pi\gamma}(x,y) 
        & = & \exp[-2.447 + 0.796x + (0.0338+0.0528x)y  
\nonumber\\      &   & + (-21.447+8.2179x)/y + (1.52436-0.38562x)/y^2],
\\
        F_{\rho \rightarrow \pi\pi\gamma}(x,y) 
        & = & \exp[-6.295 + 1.6459x + (-0.4015+0.089x)y  
\nonumber\\      &   & + (-0.954+2.05777x)/y].
\eea
No parameterization is used for the decay of the $\omega$ meson since the exact
expression requires only a one-dimensional integration~\cite{KAPUSTA_1991}
\be
        \left. E\,\frac{dN}{d^4x\,d^3p} \,\right|_{\omega \rightarrow \pi\gamma}
        \!\!\!\! = \frac{8.93\!\times\!10^{-6}\,\GeV}{E}
        \!\int^{\infty}_{E_{min}}\!\!\!dE_{\omega} 
        \, E_{\omega} \, f_{B}(E_{\omega})\left[1 + f_{B}(E_{\omega}-E)\right],
\label{omega->pigammaKAPUSTA}
\ee
where $E_{min} = 1.03\,(E^2 + 0.14\;\GeV^2)/E$ and $f_{B}(E) = 1/[\exp(E/T) - 1]$.
\end{appendix}
%

\end{document}